\documentclass[12pt]{article}
\usepackage{graphicx, epsfig, color}
\usepackage{amssymb,amsmath}

\def\eslt{E_T^{\rm miss}}
\def\delew{\Delta_{\rm EW}}

\def\delbg{\Delta_{\rm BG}}
\def\to{\rightarrow}

\def\bi{\begin{itemize}}
\def\ei{\end{itemize}}

\def\tst{\tilde t}

\def\tg{\tilde g}

\def\tw{\widetilde W}
\def\tz{\widetilde Z}
\def\alt{\lesssim}
\def\agt{\gtrsim}
\def\be{\begin{equation}}  
\def\ee{\end{equation}}  
\def\bea{\begin{eqnarray}}  
\def\eea{\end{eqnarray}}  
\def\beas{\begin{eqnarray*}}  
\def\eeas{\end{eqnarray*}}

\newcommand\prd[3]{{\it Phys.\ Rev.\ }{\bf D #1} (#2) #3}

\newcommand\prl[3]{{\it Phys.\ Rev.\ Lett.\ }{\bf #1} (#2) #3}
\newcommand\plb[3]{{\it Phys.\ Lett.\ }{\bf B #1} (#2) #3}
\newcommand\jhep[3]{{\it J. High Energy Phys.\ }{\bf #1} (#2) #3}

\newcommand\epjc[3]{{\it Eur.\ Phys.\ J. }{\bf C #1} (#2) #3}

\begin{document}
\begin{titlepage}
\begin{flushright}
UH-511-1309-20
\end{flushright}

\vspace{0.5cm}
\begin{center}
{\Large \bf  Natural Supersymmetry: Status and Prospects
}\\ 
\vspace{1.2cm} \renewcommand{\thefootnote}{\fnsymbol{footnote}}
{\large 
Xerxes Tata$^1$\footnote[1]{Email: tata@phys.hawaii.edu } 
}\\ 
\vspace{0.5cm} \renewcommand{\thefootnote}{\arabic{footnote}}
  
{\it 
$^1$Dept. of Physics and Astronomy,
University of Hawaii, Honolulu, HI 96822, USA \\
}
\end{center}

\vspace{0.5cm}
\begin{abstract}
\noindent 
\vspace*{0.3cm}

The realization that supersymmetry (SUSY), if softly broken at the weak
scale, can stabilize the Higgs sector led many authors to explore the
role it may play in particle physics. It was widely anticipated that
superpartners would reveal themselves once the TeV scale was probed in
high energy collisions. Experiments at the LHC have not yet revealed any
sign for direct production of superpartners, or for any other physics
beyond the Standard Model. This has led to some authors to question
whether weak scale SUSY has a role to play in stabilizing the Higgs sector,
and to seek alternate mechanisms for stabilizing the weak scale. We
reevaluate the early arguments that led to the expectations for light
superpartners, and show that SUSY models with just the minimal particle
content may well be consistent with LHC (and other) data and
simultaneously serve to stabilize the Higgs sector, if model parameters
generally regarded as independent turned out to be appropriately
correlated. In our view, it would be premature to ignore this
possibility, given that we do not understand the underlying mechanism of
SUSY breaking.  We advocate using the electroweak scale quantity,
$\delew$, to determine whether a given SUSY
spectrum might arise from a theory with low fine-tuning, even when the
parameters correlations mentioned above are present. We find that
(modulo technical caveats) all such models contain light higgsinos and
that this leads to the possibility of new strategies for searching for
SUSY. We discuss phenomenological implications of these models for SUSY
searches at the LHC and its luminosity and energy upgrades, as well as
at future electron-positron colliders. We conclude that natural SUSY,
defined as no worse than a part in 30 fine-tuning, will not escape
detection at a $pp$ collider operating at 27~TeV and an integrated
luminosity of 15~ab$^{-1}$, or at an electron-positron collider with a
centre-of-mass energy of 600~GeV.

\end{abstract}

\end{titlepage}

\section{Introduction} \label{sec:intro}

It has been known for a long time \cite{gildner} that the scalar sector
of the Standard Model (SM) exhibits quadratic sensitivity to the highest
mass scale ($M_{\rm high}$) in the larger theory that the SM might be
coupled to. This can be seen from the structure of
Eq.~(\ref{eq:generic}), valid in a generic quantum field theory. The
squared physical mass of a spin-zero field (such as the Higgs field of
the SM) is given in terms of the corresponding renormalized Lagrangian
parameter, $m_{\phi 0}^2$, by,
\be m_{\phi}^2 = m_{\phi 0}^2 + C_1 \frac{g^2}{16\pi^2}M_{\rm high}^2 +
C_2 \frac{g^2}{16\pi^2}m_{\rm low}^2 \log\left(\frac{M_{\rm
    high}}{m_{\rm low}}\right) +C_3 \frac{g^2}{16 \pi^2}m_{\rm
  low}^2\;.
\label{eq:generic}
\ee 
In Eq.~(\ref{eq:generic}), $m_{\rm low}$ is the highest mass scale in
the SM which is assumed to be the low energy effective theory valid
well below the energy scale $M_{\rm high}$, $g$ denotes a typical
coupling constant and the $C_i$ are dimensionless numbers typically
${\cal O}(1) \times $ spin and multiplicity factors.  The $C_3$ term
could also include ``small logarithms'' $\log(m_{\rm low}^2/m_\phi^2)$
that we have not made explicit. Well below the energy scale $M_{\rm
  high}$, the renormalizable interactions of the SM yield a good
description of nature, but at higher energies, novel effects not
present in the SM become important.

Of particular interest in particle physics are Grand Unified Theories
(GUTs) where the SM gauge group is envisioned as part of a simple group
which is spontaneously broken at a scale $M_{\rm GUT} \gg m_{\rm low}$. The
seemingly disparate SM gauge forces that we observe then arise from a
single force, and the SM gauge coupling parameters (renormalized at the
energy scale $Q\sim M_{\rm GUT}$) all assume a common value. In this case,
$M_{\rm high}$ in Eq.~(\ref{eq:generic}) is $M_{\rm GUT} \sim
10^{15-16}$~GeV in the simplest models. We then see that the $C_1$ term
in Eq.~(\ref{eq:generic}) is then $\sim 10^{28-30}$~GeV$^2$, and to
obtain the observed value of $(125~{\rm GeV})^2$ for the squared Higgs
boson mass
requires that the Lagrangian parameter $m_{\phi 0}^2$ (other terms are
much smaller) to also be as large and {\em finely tuned} to cancel
against the $C_1$ term to many significant figures. While this is
possible in principle, there is no apparent reason for this
cancellation between terms that appear to originate in different
sectors. We refer to this need for fine-tuning of seemingly unrelated
model parameters as the {\em Big Hierarchy Problem.}\footnote{For a
  contrarian philosophy, see Ref.\cite{goran}.} This problem
disappears if there are new degrees of freedom beyond those of the SM 
below the few TeV scale; {\it i.e} $M_{\rm high} \sim {\cal(O)}({\rm TeV})$.

Supersymmetry (SUSY) entered the mainstream of particle physics about
four decades ago when it was realized that supersymmetric extensions of
the SM provide an elegant solution \cite{hier} to the Big Hierarchy
problem because the $C_1$ term is absent if SUSY is softly
broken.\footnote{P. Fayet was a notable exception in that he was already
  exploring implications of SUSY for particle physics before this
  time. }  In SUSY GUT models, the $C_2$ term then dominates and, since
the large logarithm (roughly) compensates the loop factor $16\pi^2$, we
see that we would again need an unexplained cancellation between this
term and $m_{\phi 0}^2$ if $m_{\rm low}$ is significantly larger than
$m_\phi^2$. Here, $m_{\rm low}$ is again the mass scale of the heaviest
particle (with significant coupling to the Higgs boson) in the low
energy effective theory that we now use to evaluate the corrections to
the Higgs boson mass. This is, of course, no longer the SM but its
supersymmetric extension, the Minimal Supersymmetric Standard Model
(MSSM) \cite{mssm}, or one of its variants. This simple argument was the
underlying reason for the optimism in the community that at least some
SUSY partners would be found with masses ``not far above the weak
scale''.

The direct search for the superpartners, which has been one of the
central items on the agenda of $e^+e^-$, $ep$ and hadron collider
experiments at the energy frontier for well over three decades now, has
yielded no clear sign of these. Assuming a mass gap (between the parent
particle and the lighter daughter to which it decays) in excess of
several hundred GeV, various {\em simplified model analyses} by the CMS
\cite{cmsgluino} and ATLAS \cite{atlasgluino} collaborations at the LHC
have yielded lower limits on the masses of gluino and (first generation)
squarks in excess of 2~TeV. Corresponding limits on third generation
squarks exceed 1~TeV \cite{stops}. Assuming charginos (neutralinos)
decays via $\tw_1 \to W+\tz_1$ ($\tz_2 \to Z,h+\tz_1$), lower limits on
electroweak-inos of up to 600-700~GeV have been obtained for $m_{\tz_1}
< 200-300$~GeV \cite{ewinos}. In addition, low energy experiments
searching for quantum effects of supersymmetric particles that would
modify the properties of quarks and leptons; {\it e.g.} rare decays of
bottom mesons \cite{pdg} or the magnetic moment of the muon
\cite{gminus2}, have not found an unambiguous signal. Finally, searches
for (weakly interacting massive particle) dark matter, which are
frequently interpreted in the context of supersymmetric models, have
also turned up empty \cite{dmsearch}.

We should mention that in the 1980s, there were several other proposals
that attempted to address the hierachy issue. Some of these only seemed
to ``postpone the problem'' by arranging the $C_1$ contribution in
Eq.~(\ref{eq:generic}) to enter only at two loop: then the corresponding
value of $M_{\rm high}$ is an order of magnitude larger than the
simplest expectation $M_{\rm high} \sim 10 m_{\phi}\sim \cal{O}({\rm
  TeV})$, and so beyond the LHC reach. A particularly attractive
suggestion was that the Higgs scalar is really a (light) composite of
new heavy fermions, bound by new ``technicolour'' gauge forces: since
there is no elementary spin-zero field, there is no big hierarchy
problem. While this worked very well for obtaining gauge boson masses,
it led to very baroque constructions when it came to fermion masses,
consistent with absence of flavour-changing neutral currents
\cite{technicolour}. Only weak scale supersymmetry and warped extra
dimension models \cite{warped} allowed the possibility of consistently
extending the SM to very high scales.  More recently, the relaxion idea
\cite{relaxion} (also not yet realized in a UV complete model) has been
suggested, where the large hierarchy is explained through a cosmological
process that does not seem to require any precise adjustment of
parameters. A discussion of alternatives to supersymmetry for solving
the big hierarchy problem is beyond the scope of this paper.  Our
purpose here is to examine whether the non-appearance of any
superpartners in experiments at the LHC negates our primary motivation
for weak scale supersymmetry playing a role in particle physics by
stabilizing the SM Higgs sector when it is coupled to high scale
physics, as {\it e.g} in a SUSY GUT.

We emphasize that there are several other reasons for considering
supersymmetry as a key ingredient of particle physics. Ever since the
early 1980s, it has been recognized that \cite{reviews}:
\bi
\item The largest possible symmetry of the $S$-matrix includes SUSY \cite{hln};
\item Supersymmetry allows a synthesis between bosons and fermions never
  before achieved \cite{susy}; 
\item Local SUSY includes gravity \cite{local};
\item SUSY theories could include a viable candidate for (or, after what
  we have learnt now, at least a component of) dark matter \cite{gold}
  if, motivated by considerations of proton stability, we impose the
  conservation of $R$-parity.
\ei 
We stress that none of these arguments provide any indication of the
SUSY breaking scale. It is only if we require SUSY to ameliorate the
big hierarchy problem, we find that the effective SUSY breaking scale cannot be
much larger than the weak scale.  

When the gauge couplings (really speaking, the value of
$\sin^2\theta_W$) were first measured in LEP experiments, it was
recognized that these (nearly) unify in SUSY GUTs, but not in the SM
\cite{lep}. Moreover, the measured value of the Higgs boson mass
\cite{higgsmass} fits within the narrow range $m_h \alt 135$~GeV allowed
in the MSSM \cite{range}.  In extended models, assuming perturbativity
up to the GUT scale, the allowed range is not much larger \cite{larger}.

The fact that LHC experiments have led to no direct evidence for
superpartners (or for that matter any other new physics that could ease
the hierarchy problem) has led some authors to argue that because the
stop mass scale already exceeds a TeV, SUSY models already need fine
tuning at about a part per mille. This is frequently referred to as the
{\em Little Hierarchy Problem}, and has resulted in novel
(and sometimes rather complicated) proposals for its resolution. While new
theoretical ideas are obviously always welcome, one of our goals is to
critically assess whether LHC data indeed imply the existence of a
little hierarchy that calls for the abandonment of the simple, calculable
(and hence predictive) picture of perturbative SUSY GUTs.\footnote{We
  stress that SUSY clearly provides a solution to the Big Hierarchy
  problem as long as $M_{\rm SUSY} \ll M_{\rm GUT}$. We leave it to the
  reader to examine whether the proposed alternatives to SUSY truly
  address the hierarchy problem beyond leading loop order, and if they do, to
  assess the pros and cons of the new proposals over SUSY GUTs.}

\section{The Mass Scale of Superpartners} \label{sec:scale}

Let us start by recalling why it was that superpartners were expected to
be close to the weak scale.
In SUSY GUTs, since the logarithm in Eq.~(\ref{eq:generic}) is about 30,
the leading correction,
$$\frac{\delta m_h^2}{m_h^2} \sim C_2 \frac{g^2}{16\pi^2} \frac
{m_{\rm SUSY}^2}{m_h^2} \log\left(\frac{M_{\rm GUT}}{m_{\rm
    low}}\right),$$ rapidly exceeds unity if $m_{\rm SUSY}$ is
significantly larger than $m_h$. Many authors argued that in order not
to have unexplained cancellations, it is reasonable to set $\delta
m_h^2 \lesssim m_h^2$, and , $\Delta_{\rm log} = {\delta m_h^2 \over
  m_h^2}$ was suggested \cite{knpap} as a simple measure of the degree
of fine-tuning, and continues to be used by some authors. What went
wrong?
\bi
\item Perhaps, $\delta m_h^2 < m_h^2$ is too stringent a requirement; we
  know many examples of accidental cancellations of an order of
  magnitude. While an unexplained cancellation of two orders of
  magnitude is, perhaps, too strong, accidental cancellations of an
  order of magnitude are not uncommon. The well known factor of
  $\pi^2-9$ in the decay rate of orthopositronium provides an
  ``accidental cancellation'' of an order of magnitude. While this is
  somewhat subjective, we will draw the line halfway in between, and
  require unexplained cancellations to be smaller than a part in
  30.\footnote{Amusingly, the angular sizes of the sun and moon (viewed
    from earth) are the same to within this precision, another example
    of an accident.}

\item These ``naturalness bounds'' 
  apply only to those superpartners with large couplings to the Higgs
  sector, and so do not apply to first (or even second generation)
  squarks and gluinos whose masses are most stringently probed at the
  LHC. These superpartners couple to the Higgs sector only at two-loop
  so that their masses could easily be $\sim 5-10$~TeV or more
  because there would be an additional $16\pi^2$ in the $C_2$ term of
  Eq.~(\ref{eq:generic}).\footnote{The
  $D$-term coupling contributions largely cancel.}

\item There are various one-loop contributions to the $C_2$ terms in
  Eq.~(\ref{eq:generic}) that could, in principle, cancel against one
  another.  Using $\Delta_{\rm log}$ as a measure of the degree of
  cancellations {\em assumes that  contributions from various
    superpartners are all independent}. However, since we all
  expect that various superpartner masses will be correlated once we
  understand the mechanism of supersymmetry breaking,  {\em
    automatic cancellations} between contributions from various
  superpartners could well occur when we evaluate the fine-tuning in any
  high scale theory. {\it Ignoring these correlations, will overestimate
    the ultra-violet sensitivity of any model.}  \ei

Parameter correlations
  are most simply incorporated into the most commonly used fine-tuning
  measure introduced by Ellis, Enqvist, Nanopoulos and Zwirner
  \cite{eenz} and subsequently explored by Barbieri and Guidice
  \cite{bg}: 
\be 
\Delta_{\rm BG}
\equiv max_i\left|\frac{p_i}{M_Z^2}\frac{\partial M_Z^2}{\partial
  p_i}\right|\;.
\label{eq:DBG}
\ee
Here, the value of $M_Z^2$ is a prediction in terms of $p_i$'s, the {\em
  independent} underlying parameters of the theory.  It does not matter
that $M_Z^2$ rather than $m_h^2$ is used to define the sensitivity
measure since both the quantities are proportional to the
square of the Higgs field $vev$. The important point is that
$\delbg$ here measures the sensitivity with respect to the {\em
  independent} parameters of any model and so takes into account the
correlations that we mentioned. Since $\delbg$ ``knows about''
correlations that are ignored in $\Delta_{\rm log}$, we expect
$\Delta_{\rm log} \ge \delbg$, which is why we said earlier that
$\Delta_{\rm log}$ would over-estimate the degree of fine-tuning.

The problem, of course, is that without a detailed knowledge of how
superpartners acquire their masses, it is not possible to evaluate how
these correlations affect the UV-sensitivity. We will see in
Sec.~\ref{subsec:ewft} that we can, however, obtain a robust lower bound
on $\delbg > \delew$, where $\delew$ is determined only by the {\em weak
  scale SUSY parameters} which (in principle) can be directly measured
if superpartners are discovered. In line with our earlier discussion,
models with $\delew > 30$ can then unambiguously be regarded as fine-tuned.

\subsection{Electroweak Fine-tuning: A lower limit on $\delbg$} \label{subsec:ewft}

The value of $M_Z^2$ obtained from the minimization of the
one-loop-corrected Higgs boson potential of the MSSM,
\be 
\frac{M_Z^2}{2} = \frac{(m_{H_d}^2+\Sigma_d^d) - (m_{H_u}^2+\Sigma_u^u) \tan^2\beta}{\tan^2\beta -1} -\mu^2,
\label{eq:mZsSig}
\ee 
is our starting point.  Eq.~(\ref{eq:mZsSig}) is obtained using the weak
scale MSSM Higgs potential, with all parameters evaluated at the scale
$Q=M_{\rm SUSY}$.  The $\Sigma$s in Eq.~(\ref{eq:mZsSig}), which arise
from one loop corrections to the Higgs potential, are analogous to the
$C_3$ term in (\ref{eq:generic}). Explicit forms for the $\Sigma_u^u$
and $\Sigma_d^d$ may be found in the Appendix of Ref.~\cite{rns}.

We require that the observed value of $M_Z^2$ is obtained without large
cancellations between terms on the right-hand-side of
Eq.~(\ref{eq:mZsSig}),
{\it i.e} none of these terms are hierarchically  larger
than $M_Z^2$.  Electroweak fine-tuning of
$M_Z^2$ can then be quantified by
\cite{ltr,baersugra,rns},
 \be \Delta_{\rm EW} \equiv max_i
\left|C_i\right|/(M_Z^2/2)\;.
\label{eq:delew}
\ee 
Here, $C_{H_d}=m_{H_d}^2/(\tan^2\beta -1)$,
$C_{H_u}=-m_{H_u}^2\tan^2\beta /(\tan^2\beta -1)$ and $C_\mu =-\mu^2$.
Also, $C_{\Sigma_u^u(k)} =-\Sigma_u^u(k)\tan^2\beta /(\tan^2\beta -1)$
and $C_{\Sigma_d^d(k)}=\Sigma_d^d(k)/(\tan^2\beta -1)$, where $k$ labels
the various loop contributions included in Eq.~(\ref{eq:mZsSig}).  We
immediately see that any upper bound on $\delew$ that we impose from
electroweak naturalness considerations  implies a
corresponding limit on $\mu^2$, a connection first noted two
decades ago \cite{CCN}. 

Since $|\mu|$ sets the scale for the doublet higgsino mass, we are led
to infer that these {\em higgsinos cannot be hierarchically heavier than
  $M_Z$} in any theory with small values of $\delew$. There are,
however, potental loopholes that could void this conclusion that we make
explicit.
\bi
\item We have implicitly assumed that the superpotential parameter $\mu$
  is independent of the soft SUSY breaking (SSB) parameters. If $\mu$
  were correlated to the SSB parameters -- in particular with
  $m_{H_u}^2$ -- there could be automatic cancellations that would
  preclude us from concluding that higgsinos are light. Put differently,
  we assume that the superpotential and SSB breaking sectors could have
  different physical origins, and so are unrelated.

\item We assume that there is no SSB contribution to the
  higgsino mass and that the $\mu^2$ that enters in
  Eq.~(\ref{eq:mZsSig}) via the scalar Higgs potential is also the
  higgsino mass parameter. Such a term would break SUSY softly as long
  as there are no SM singlets with significant couplings to the
  higgsinos \cite{ross}.
We note that Nelson and Roy
  \cite{nr} and Martin \cite{martin} have constructed models with
  additional adjoint chiral superfields at the weak scale in which the
  SUSY mass parameters in the Higgs boson sector are logically independent of
  higgsino masses.

\item It has been pointed out \cite{ckl} that if the Higgs particle is a
  pseudo-Goldstone boson in a theory with an almost exact global
  symmetry, it is possible that the Higgs boson remains light even if
  the higgsinos are heavy because cancellations that lead to a low Higgs
  mass (and concomitantly low $M_Z^2$) are a result of a symmetry. Such
  models necessarily include additional fields in order to have complete
  multiplets of the global symmetry.

\ei

Despite these exceptions (all of which require the introduction of
new low energy fields that serve no other purpose), we find it
compelling that in models with a minimal (low energy) particle content
the higgsino mass enters Eq.~(\ref{eq:mZsSig}) directly, so that a low
value of $\delew$ implies the existence of doublet higgsinos with masses
not far above $M_Z$. We see no  phenomenological motivation for the
introduction of these extra fields at the weak scale, and so will
continue to regard the existence of light higgsinos as a robust
phenomenological consequence of natural SUSY in the rest of this paper.

The requirement of electroweak naturalness imposes {\em upper limits} on other
superpartner masses, over and above the higgsino limit that we have just
discussed. We will see below that models with stops as heavy as 3.5~TeV
and gluinos as heavy as 6~TeV can be compatible with $\delew < 30$, in
sharp contrast to stop bounds in the few hundred GeV range that
emerge\footnote{Recall that we saw in Sec.~\ref{sec:intro} that this was
  the cause for disenchantment with SUSY in some quarters.} if the
possibility of parameter correlations is ignored. First and second
generation sfermions can be as heavy as tens of TeV, provided the
sfermion spectrum exhibits well-motivated (partial) degeneracy patterns
\cite{maren}. These heavy sfermions then ameliorate SUSY flavour and $CP$
problems.

We note here that $\delew$ as defined here entails only weak scale
parameters and so has no information about the $M_{\rm high}$ terms that
cause weak scale physics to exhibit logarithmic sensitivity to high
scale physics.  For this reason, $\delew$ does not measure the UV
sensitivity of the underlying high scale theory, as already noted in
Ref.~\cite{rns}. However, precisely because $\delew$ does not contain
information about the large logs, we expect (modulo technical
caveats that we will not get into here \cite{am}) that $$\delew \leq
\delbg.$$ Thus $\delew^{-1}$ is the minimum fine-tuning in any theory
with a given superpartner spectrum, as noted just before the start of
Sec.~\ref{subsec:ewft}.

Although it is not a fine-tuning measure of a high scale theory,
$\delew$ is nonetheless useful for many reasons.
\bi

\item Since it depends only on weak scale parameters, $\delew$ is
  essentially determined by the SUSY spectrum, and so is
  ``measureable'', at least in principle. 

\item If $\delew$ turns out to be large, the underlying theory that
  yields this spectrum will be fine-tuned because $\delbg$ is even
  larger. While small $\delew$ does not necessarily imply the absence
  of fine-tuning, it leaves open the possibility of finding an
  underlying natural theory with the same superpartner spectrum where
  SSB parameters are correlated so that the large logarithms in the
  $C_2$ term of Eq.~(\ref{eq:generic}) nearly cancel.\footnote{The
    possibility that correlations among underlying parameter reduces
    the fine-tuning has been noted by other authors \cite{reduce}.}
  In a top-down theory which has such correlations among the SSB
  parameters, $\delbg$ will be numerically close to $\delew$. Sec. 3
  of Ref.~\cite{am} illustrates how the cancellations might occur.  

\item In the absence of a complete understanding of how superpartners
  acquire masses and SUSY breaking couplings, it is not possible to
  evaluate $\delbg$ with all the parameter correlations correctly
  incorporated. We advocate instead that $\delew$ be used for any
  discussion of fine-tuning because, though it may {\em underestimate}
  the degree of fine-tuning, it at least allows for the possibility that
  SUSY parameters frequently taken to be independent may turn out to be
  correlated. Disregarding this possibility may cause us to discard
  otherwise perfectly viable phenomenological models \cite{bbm1,am}. We
  note that $\delbg$ naively computed {\it i.e.} without parameter
  correlations included, could well be two orders of magnitude larger
  than $\delew$ \cite{am}.

\item Broad aspects of SUSY phenomenology are determined by the
  superpartner spectrum. An investigation of the phenomenology of models
  with low $\delew$ is, therefore, in effect an investigation of the
  phenomenology of the underlying (potentially) natural underlying theories.
\ei

\section{Models with Low $\delew$}
\label{sec:spectra}

We have seen that the the magnitude of $\mu$ is fixed
using Eq.~(\ref{eq:mZsSig}) which
is well approximated by,
$$
\frac{1}{2}M_Z^2 \simeq -(m_{H_u}^2+\Sigma_u^u) -\mu^2,
$$ 
for moderate to large values of $\tan\beta$. Except for radiative
corrections, a weak scale value of $-m_{H_u}^2$ close to $M_Z^2$ ensures
a comparable value of $\mu^2$, so that $\delew$ is also not far above
unity.  This is, however, a non-trivial constraint on $m_{H_u}^2$ that cannot
always be consistently realized.  Within the much-studied mSUGRA
framework \cite{msugra} $m_{H_u}^2$ evolves to a negative value at the weak scale
(this is the celebrated mechanism of radiative electroweak symmetry
breaking \cite{radewsb}), and its magnitude is comparable to that of
other weak scale SSB parameters.  The resulting value of $\mu^2$ is --
taking experimental constraints on sparticle masses into account --
typically much larger than $M_Z^2$ as long as the radiative corrections
contained in $\Sigma_u^u$ are of modest size. Indeed, within the mSUGRA
framework, one cannot obtain $\delew \alt 100$ consistently with the
observed value of $m_h$ \cite{baersugra}.

A small weak scale value of $m_{H_u}^2$ can always be obtained if we
relax the assumption of high-scale scalar mass parameter universality
that is the hallmark of mSUGRA, and treat the Higgs field mass
parameters as independent of corresponding matter scalar masses. The
Non-Universal Higgs Mass model, which has two additional GUT scale
parameters $m_{H_u}^2$ and $m_{H_d}^2$ (NUHM2 model) over and above
the the mSUGRA parameter set: $m_0, m_{1/2}, A_0, \tan\beta, \ {\rm
  and} \ sign(\mu)$, provides an appropriate setting \cite{nuhm2}.  It
is also worth remarking that the large value of the trilinear third
generation SSB scalar coupling required to obtain low
values of $\delew$ simultaneously raises the Higgs boson mass to its
observed value \cite{ltr}.  The NUHM3 model where third generation
sfermion mass parameter is independent of $m_0$ as well as the SUSY
breaking Higgs boson mass parameters provides an even more general
parametrization for phenomenological analyses. There are some top-down
scenarios in which this splitting of the third generation mass
parameter is expected.

In these NUHM frameworks, the three MSSM gaugino masses are assumed to
arise from a single gaugino mass parameter (renormalized at $Q=M_{\rm
  GUT}$) in the same way the SM gauge couplings arise from a single
unified gauge coupling in SUSY GUTs. While this appears to be well
motivated, it is important to recognize that gaugino mass unification --
unlike the unification of gauge couplings -- is not compulsory even in SUSY
GUTs: {\em tree level} gaugino mass parameters, renormalized at the
appropriate high scale, unify only if the field that breaks SUSY is a
singlet of the GUT group \cite{unifgaug}. Radiative corrections
evaluated by the renormalization group evolution of gaugino mass
parameters from $Q=M_{\rm GUT}$ to the sparticle mass scale results in
the familiar pattern of weak scale gaugino mass parameters: $m_{\tg}
\simeq 3M_2 \simeq 6M_1$, resulting in relatively large mass splittings
between the spin-${1\over 2}$ SUSY partners of the gauge bosons. Very
different mass patterns, and correspondingly different phenomenology,
may be possible  if gaugino mass unification if gaugino
mass unification is not assumed.

Non-universal gaugino mass patterns are also possible if gaugino masses
arise only at the loop level.  In supergravity models, there is a loop
contribution to gaugino (and other superpartner) masses that arises from
a breaking of scale invariance induced by quantum anomalies. This
anomaly contribution to gaugino masses, proportional to the
corresponding gauge $\beta$-function, is always present but because it
is suppressed by a loop factor is important only if the tree-level
contributions are absent or strongly suppressed. This happens in the
so-called anomaly mediated SUSY breaking (AMSB) models \cite{amsb} and
their variants.

It is not our purpose here to go into the pros and cons of various SUSY
models. Since our goal is to explore the phenomenology of natural SUSY
models, we confine ourselves to the study of the variety of
spectra and phenomenology that may be possible in various well-motivated
natural SUSY frameworks that allow $\delew < 30$, consistently with current
experimental constraints. Models that we consider
include: 
\bi
\item natural NUHM2 and NUHM3 (hereafter denoted by nNUHM2 and
  nNUHM3) models that we adopt as representative of 
models with gaugino mass unification at the GUT scale;

\item a phenomenological generalization \cite{namsb} of the AMSB
  framework \cite{amsb} with non-universal bulk Higgs mass parameters
  and trilinear couplings to allow $m_h\simeq 125$~GeV with $\delew <
  30$ (nAMSB). The gaugino mass pattern is as given by AMSB discussed
  above, and very different from the pattern expected in models with
  gaugino mass universality.

\item a phenomenological generalization \cite{ngmm} of the original
  mirage-mediation framework \cite{mirage} in which one expects
  comparable gravity-mediated and anomaly-mediated contributions to SSB masses and couplings, allowing for patterns of SUSY
  spectra not realizeable in other frameworks. The hallmark of this
  class of natural generalized mirage models (nGMM) is that gaugino mass
  parameters apparently (almost) unify at a scale $Q=\mu_{\rm mirage}$,
  determined by the relative value of gravity and anomaly-mediated
  contributions to the SSB parameters.  In particular, if
  $\mu_{\rm mirage}$ is not far above the sparticle mass scale, the (low
  energy) gaugino mass parameters only have small splittings resulting
  in very different phenomenology from the other scenarios. We stress
  there is no physical threshold at $Q=\mu_{\rm mirage}$, and the
  gaugino mass as well as other SUSY parameters continue to evolve
  smoothly through the mirage-unification scale all the way up to
  $M_{\rm GUT}$. The nGMM pattern of gaugino masses is also expected to
  arise in the so-called mini-landscape picture \cite{miniland} which
  targets the region of the string landscape that leads to the MSSM as
  the low energy effective theory. The phenomenology of the natural
  string mini-landscape picture is studied in Ref.\cite{minilandBaer}.
  \ei

Each of these frameworks allow spectra with $\delew < 30$, consistently
with all experimental constraints.  In the following, use these models
to guide our exploration of the phenomenlogy of natural SUSY.  We will
adopt the NUHM models as typifying natural SUSY models with gaugino mass
unification, while the nAMSB and nGMM models allow the exploration of
natural SUSY where gaugino mass patterns deviate from their universal
values in well-motivated ways. The nGMM model can accommodate a
compressed as well as very split gaugino mass spectrum.

\section{Phenomenology} \label{sec:phen}

As already emphasized, charged and neutral higgsinos with masses ranging
from about 100~GeV (to evade LEP2 limits) to 300-350~GeV (so that
$\delew < 30$) are the hallmark of all natural SUSY models. In models
with gaugino mass unification typified by nNUHM2, nNUHM3 models, the
heavier charged and neutral higgsinos have a mass gap of 10-30 GeV with
the lightest supersymmetric particle (LSP) that escapes detection at
particle accelerators. Smaller mass gaps are possible in natural SUSY
only if we give up gaugino mass unification. Other superpartners may be
much heavier even with $\delew < 30$ as we have already mentioned.  
Here, we present an overview of various SUSY signals in natural SUSY
scenarios, with an emphasis on signatures suggestive of light higgsinos
in the spectrum.

\subsection{LHC and its luminosity upgrade} 
\label{subsec:lhc}

\begin{figure}[tbh]{\begin{center}
\includegraphics[width=12cm,clip]{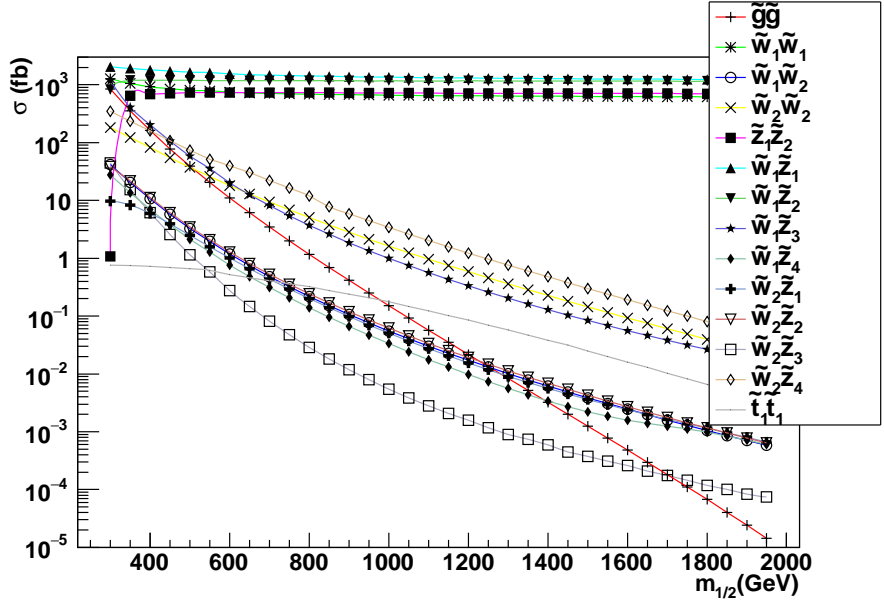}
\caption{Various NLO sparticle pair production cross sections
versus $m_{1/2}$ along the NUHM2 model line (\ref{eq:mline}) for $pp$
collisions at $\sqrt{s}=14$~TeV. Results are insensitive to the choice
of $m_0$ as long as squarks are decoupled from LHC physics. }
\label{fig:csec}\end{center}}
\end{figure}

In natural SUSY the light higgsinos are likely to be the most copiously
produced superpartners at the LHC \cite{world,rnslhc}. This is
illustrated in Fig.~\ref{fig:csec} where we show various -ino production
cross sections versus $m_{1/2}$, for the NUHM2 model-line with
\be
m_0=5~{\rm TeV}, A_0=-1.6m_0, \tan\beta=15, \mu=150~{\rm GeV}, \ {\rm
  and} \  m_A=1~{\rm TeV},
\label{eq:mline}
\ee
at LHC14. We have traded the high scale values of the Higgs mass
parameters in favour of $\mu$ and $M_A$. Our choice of $m_0$ ensures
that squarks are heavy so that we have agreement with flavour
constraints. Note that the low $m_{1/2}$ portion of the graph is
excluded by LHC constraints. 

The cross sections for the production of higgsino-like charginos and
neutralinos ($\tw_1$, $\tz_{1,2}$) whose masses remain fixed close to
$|\mu| = 150$~GeV across most of the plot remain flat, while cross
sections for the gaugino-like states ($\tw_2$, $\tz_{3,4}$) fall off
because their masses increase with $m_{1/2}$. Cross sections for
gaugino-higgsino pair production are dynamically suppressed. Pair
production of gluinos and top squarks also occurs at observable rates if
these particles are kinematically accessible, while other squarks and
sleptons are essentially decoupled at the LHC.

\subsubsection{Electroweak Higgsino Pair Production} 

The small visible energy release in their decays makes signals from higgsino
pair production impossible to detect over SM backgrounds. We are thus led
to investigate other strategies for discovery of SUSY.

\subsubsection{Mono-jet and Mono-photon Signals} \label{subsec:mono}

Many groups have suggested that experiments at the LHC may be able to
identify the pair production of LSPs via high $E_T$ mono-jet or
mono-photon plus $\eslt$ events, where the jet or the photon arises
from QCD or QED radiation. A careful study of this signal for the case of
light higgsinos, incorporating the correct matrix elements for all
relevant higgsino pair production processes 
shows that it will be very difficult to
extract the signal unless SM backgrounds can be controlled at the
better than the percent level \cite{mono}. The problem is that the
jet/photon $E_T$ distribution as well as the $\eslt$ distribution has
essentially the same shape for the signal and the background.

In Ref.~\cite{ghww} it was suggested that it may be possible to enhance
the mono-jet signal relative to background by requiring additional soft
leptons in events triggered by a hard mono-jet. Ref.~\cite{kribs}
examined the mono-jet signal requiring, in addition, two opposite-sign
leptons in each event, and showed that the SUSY signal could indeed be
observable at the LHC.  Subsequent detailed studies (within the NUHM2
framework) of mono-jet, and also mono-photon, events with opposite-sign,
same-flavour dileptons with low invariant mass showed that experiments
at LHC14 would be able to detect a $5\sigma$ signal from higgsino pair
production for $|\mu|< 170$~(200)~GeV, assuming an integrated luminosity
of 300 (1000)~fb$^{-1}$ \cite{dilep}.  Very interestingly, the ATLAS
collaboration \cite{atlasdilep} has already excluded higgsino mass
values well beyond the LEP2 limits even if $m_{\tz_2}-m_{\tz_1}$ is as
small as 4~GeV, but the excluded $m_{\tz_2}$ range is very sensitive to
the mass difference, and falls rapidly once $m_{\tz_2}-m_{\tz_1}<
5$~GeV.  CMS projections \cite{cmsdilep3} for 3~ab$^{-1}$ suggest a
5$\sigma$ reach up to $\mu =240$~GeV, for $m_{\tz_2}-m_{\tz_1}\simeq
10$~GeV, while the corresponding 95\% CL exclusion extends to
350~GeV. The ATLAS 95\%CL exclusion region \cite{atlasdilep3} also
extends to 350~GeV even for $m_{\tz_2}-m_{\tz_1} \sim 4-5$~GeV, but
falls rapidly for smaller mass differences. Keeping in mind that the
higgsinos of natural SUSY may be as heavy as 300-350~GeV, we conclude
that while LHC experiments will be sensitive to the most promising part
of the parameter of natural SUSY models, they may not be able to probe
the entire natural SUSY region with $\delew \leq 30$ at the 5$\sigma$
level, especially if higgsino mass gap is significantly smaller than
$\sim 10$~GeV. The ultimate reach will depend on the degree to which the
LHC experiments will be able to reliably identify and measure
soft-leptons in events triggered by a monojet or, perhaps, a
mono-photon.  These  are channels worth watching.

\subsubsection{Same Sign Dibosons} 

Typical natural SUSY scenarios suggest that $|\mu| \ll M_{1,2}$ so that
$\tw_1$ and $\tz_2$ are higgsino-like and, in models with gaugino mass
unification, only 10-30~GeV heavier than $\tz_1$. Then $\tz_3$ is dominantly
a bino, and $\tw_2$ and $\tz_4$ are winos. For heavy squarks,
electroweak production of the bino-like $\tz_3$ is dynamically
suppressed since $SU(2)\times U(1)_Y$ symmetry precludes a coupling of
the bino to the $W$ and $Z$ bosons. However, winos have large ``weak
iso-vector'' couplings to the vector bosons so that wino pair production
occurs at substantial rates. Indeed we see from Fig.~\ref{fig:csec} that
for high values of $m_{1/2}$ the kinematically disfavoured
$\tw_2^\pm\tw_2^\mp$ and $\tw_2\tz_4$ processes are the dominant
sparticle production mechanisms\footnote{Although we  use the NUHM2
  framework for the illustration of the signal, wino pair production
  would also be possible in other models. Keep in mind though that in
  models where gaugino mass parameters do not unify at the GUT scale,
  the neutral wino could be $\tz_3$ rather than $\tz_4$.} with large
visible energy release and high $\eslt$.

Wino
production leads to a novel signature involving same-sign dibosons
%
%
%
produced via the process, $pp \to \tw_2^{\pm} (\to
W^{\pm}\tz_{1,2})+\tz_4 (\to W^\pm\tw_1^\mp)$. The visible decay
products of $\tw_1$ and $\tz_2$ tend to be soft, so that the signal of
interest is a pair of same sign, high $p_T$ leptons from the decays of
the $W$-bosons, {\em with limited jet activity in the event}
\cite{ssdb}. This latter feature serves to distinguish the wino pair
production signal from same sign dilepton events that might arise at the
LHC from Majorana gluino pair production \cite{glssl} that always has
very hard jets from the primary decay of the gluinos.  We note also that
$pp \to \tw_2^\pm \tw_2^\mp$ production (where one chargino decays to
$W$ and the other to a $Z$) also makes a non-negligible contribution to
the $\ell^\pm\ell^\pm +\eslt$ channel when the third lepton fails to be
detected. The same sign dilepton signal with limited jet activity is a
hallmark of all low $\mu$ models, as long as wino pair production is not
kinematically suppressed.

The extraction of the same sign dilepton signal from wino production
requires a detailed analysis to separate the signal from SM backgrounds:
see Sec.~5 of Ref.~\cite{rnslhc}, and also Ref.\cite{aspects} where the
analysis was re-examined and refined. The most important cuts necessary
for suppressing backgrounds are a hard cut on $\eslt$, together with a
cut on
\[
m_T^{\rm min} \equiv {\rm
  min}\left[m_T(\ell_1,\eslt), m_T(\ell_2,\eslt)\right],
\] 
along with requiring at most one jet (not tagged as a $b$-jet) in the
event.  It was shown that, with 3~ab$^{-1}$, LHC experiments would allow
a 5$\sigma$ discovery of winos with a mass up to 900~GeV. By itself, this
falls well short of the entire natural SUSY parameter space.

\subsubsection{Gluinos and stops} \label{subsubsec:gst}
\begin{figure}[tbp]
\begin{center}
\includegraphics[width=0.45\textwidth]{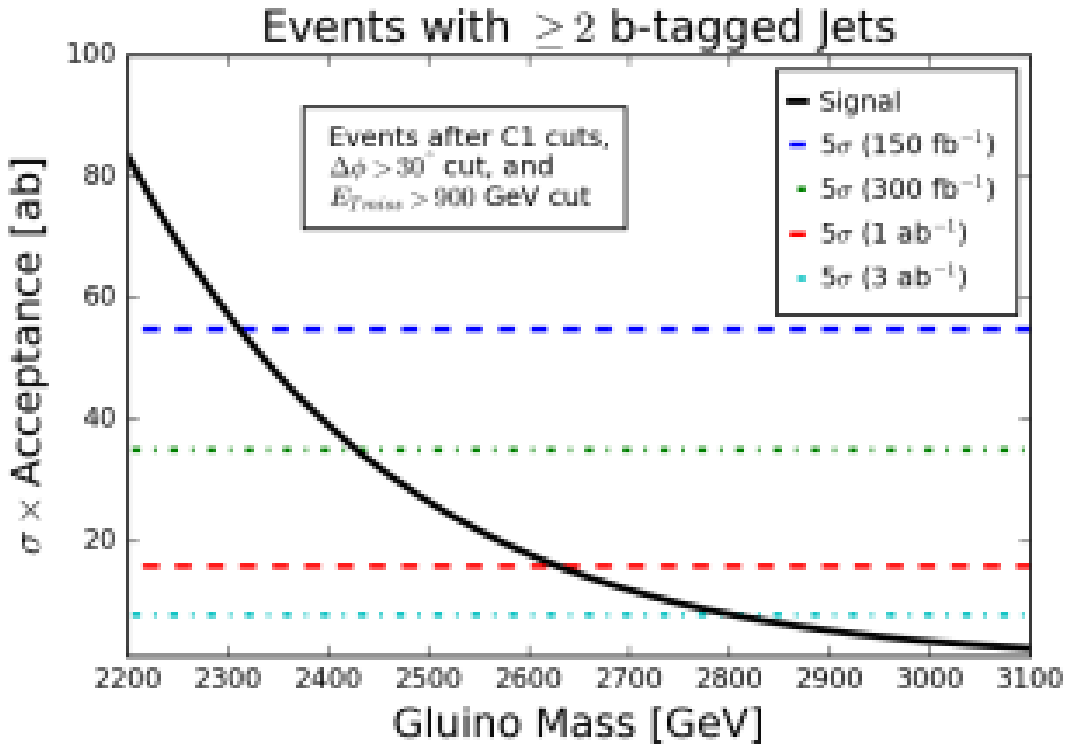}
\includegraphics[width=0.45\textwidth]{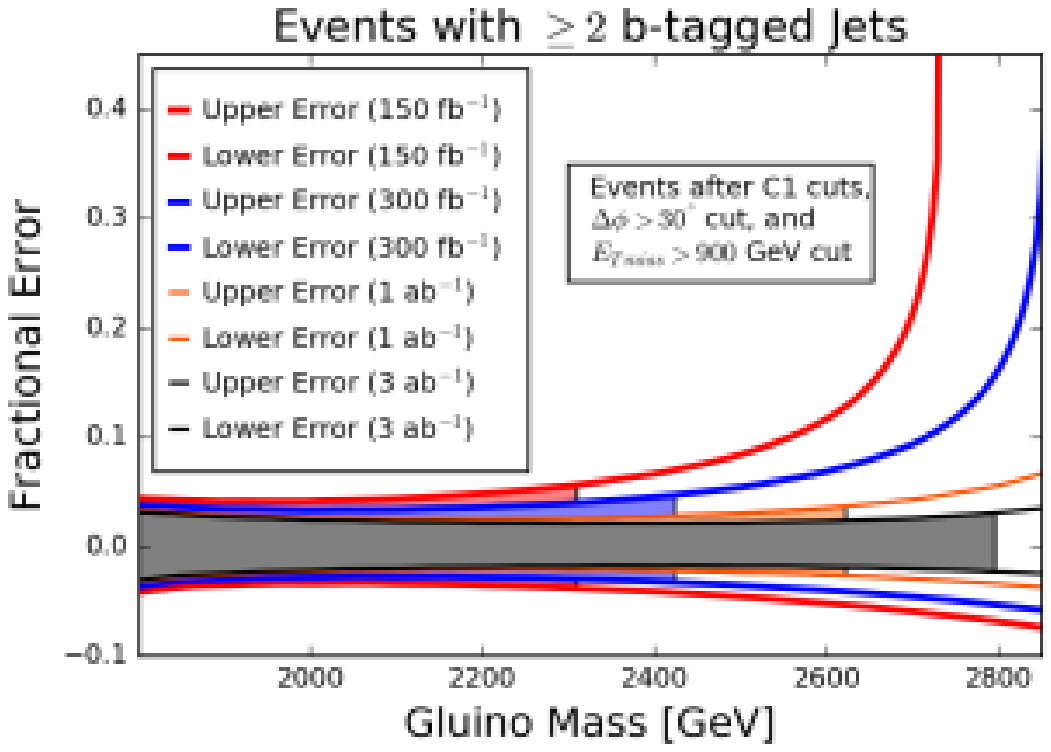}
\caption{ The left-hand frame shows gluino signal cross section for the
  $\ge 2$ tagged $b$-jet events after hard cuts detailed in
  Ref.\cite{jamielhc}. The horizontal lines show the minimum cross
  section for which the signal will be detectable with an equivalent
  Gaussian probability corresponding to $5\sigma$ above Poisson
  fluctuations of the background. The right frame shows the precision
  with which the gluino mass may be extracted from the measured rate for 
gluino events (assuming a 15\% uncertainty in the gluino cross section)
for different values of integrated luminosity at the LHC.}

\label{fig:gluino}
\end{center}
\end{figure}

Unless their production is kinematically suppressed, coloured
particles are expected to be the copiously super-partners produced at
hadron colliders.  Within natural SUSY, the lighter stop is
significantly lighter than other squarks, so that gluinos dominantly
decay via $\tg \to t\tst_1^*, \bar{t}\tst_1$, where the (real or
virtual) stop decays dominantly to higgsinos via $\tst_1\to
t\tz_{1,2}$ or $\tst_1 \to b\tw_1$. Gluino pair production is,
therefore, signalled by events with up to four hard tagged $b$-jets
and large $\eslt$. It is has been shown that it is possible to isolate
an almost pure signal sample from gluinos requiring at least four hard
jets, at least two of which are tagged as $b$-jets, and a very stiff
$\eslt$ (along with other cuts) to nearly eliminate Standard Model
backgrounds \cite{jamielhc}. With these cuts, experiments at the LHC
should be able to observe a $5\sigma$ gluino signal if $m_{\tg} < 2400
\ (2800)$~GeV for an integrated luminosity of 300 (3000)~ab$^{-1}$ in
both the two and three tagged $b$-jet channels. This is illustrated in
the left frame of Fig.~\ref{fig:gluino} for two tagged $b$-jet
events. A similar reach is obtained in the three tagged $b$-jet
channel. Unfortunately, however, this only covers part of the range of
$m_{\tg}$ allowed by natural SUSY. If, however, the gluino signal is
observable, a measurement of the rate of gluino events in the clean
SUSY sample obtained above also allows for a determination of the
gluino mass with a precision of 2.5-5\%, depending on
the integrated luminosity that is accumulated and the value of
$m_{\tg}$: see the right frame of Fig.~\ref{fig:gluino} \cite{jamielhc}.

Stop pair production occurs at a rate shown in Fig.~\ref{fig:csec} for
the NUHM2 model line introduced earlier. However, even with
3~ab$^{-1}$, the 5$\sigma$ LHC reach, assuming that $\tst_1 \to
t\tz_1$, extends out to about 1.3~TeV for $m_{\tz_1}\alt 400$~GeV,
while the 95\% CL sensitivity region extends to 1.6-1.7~TeV
\cite{atlasstophl}. Since the stop of natural SUSY dominantly decays
via $\tst_1\to t\tz_{1,2}$ or $b\tw_1$ (where $m_{\tw_1}\simeq
m_{\tz_2} \simeq m_{\tz_1}$), and the decay products of the heavier
higgsinos are essentially invisible, we expect that the natural SUSY
reach of the stop is qualitatively to similar to the numbers quoted
above. It is thus entirely possible that the stop may evade detection
at the high-luminosity LHC even in models with $\delew < 30$. Here, we
sharply differ from those authors that neglect the possibility
parameter correlations, and so conclude that the absence of any sign
of the stop would imply that SUSY is fine-tuned to parts per mille, or
worse.

\subsubsection{Other Signals} 

The hard trilepton signal from wino pair production, {\it i.e.} from
the reaction $pp \to \tw_2 \tz_4 + X \to W+Z +\eslt+ X$ in low $|\mu|$
models with gaugino mass unification, has long considered to be the
golden mode for SUSY searches~\cite{trilep}.  The leptons come from
the decays of the vector bosons, while the $\eslt$ dominantly arises
from the $\tw_1/\tz_{1,2}$ (whose visible decay products are very
soft) daughters of the winos and from the neutrino from $W$ decay. A
detailed analysis \cite{rnslhc} shows that the LHC14 reach in the
NUHM2 model extends to $m_{1/2} = 500$ (630)~GeV for an integrated
luminosity of 300 (1000)~fb$^{-1}$. This is considerably lower than
the reach via the SSdB channel, but can yield a confirmatory
signal. Much of this region has already been probed at the LHC
\cite{ewinos} albeit in simplified models.

In models with light higgsinos, the (heavy) charged wino decays via
$\tw_2\to \tz_{1,2}W$, $\tw_2\to \tw_1 Z$ or $\tw_2 \to \tw_1 h$ with
branching ratios $\sim 2:1:1$, while the neutral wino decays via $\tz_4 \to
\tw_1^{\pm}W^{\mp}$, $\tz_4\to \tz_{1,2}Z$ or $\tz_4 \to \tz_{1,2}h$
with branching ratios $\sim 2:1:1$ \cite{aspects}. Since the daughter higgsinos
are essentially invisible, wino pair production potentially leads to a
variety of interesting $VV+\eslt$ ($V=W,Z$), $Vh+\eslt$ and $hh+\eslt$
events {\em in predicted proportion.} Observation of these
events in the expected ratios would point to a model with light
higgsinos, though this may be more relevant at the proposed energy
upgrade of the LHC discussed below.

The LHC reach in the 4 lepton signal channel was also examined in
Ref.\cite{rnslhc} by requiring four isolated leptons with $p_T(\ell) >
10$~GeV, a $b$-jet veto (to reduce backgrounds from top quarks), and
$\eslt > \eslt({\rm cut})$. Potential backgrounds come from $ZZ,
t\bar{t}Z, ZWW, ZZW, ZZZ$ and $Zh(\to WW^*)$ production. It was found
that in low $|\mu|$ models, the LHC reach via the $4\ell$ search
extends somewhat beyond that in the trilepton channel.

\subsubsection{A Recap of the Reach of the LHC and its Luminosity Upgrade} 

\begin{figure}[tbp]
\begin{center}
\includegraphics[width=0.34\textwidth]{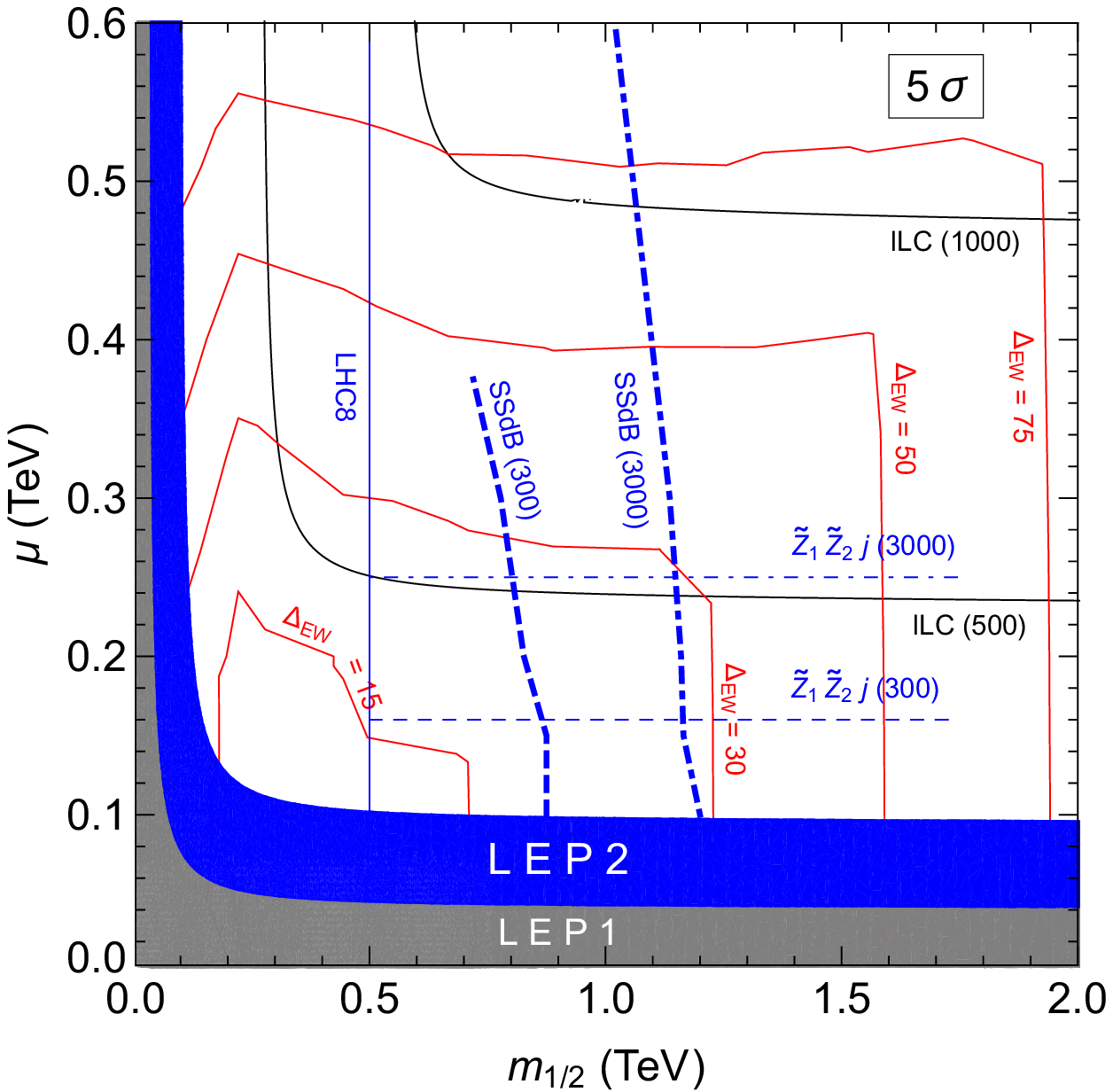}
\vspace{0.7in}
\includegraphics[width=0.51\textwidth]{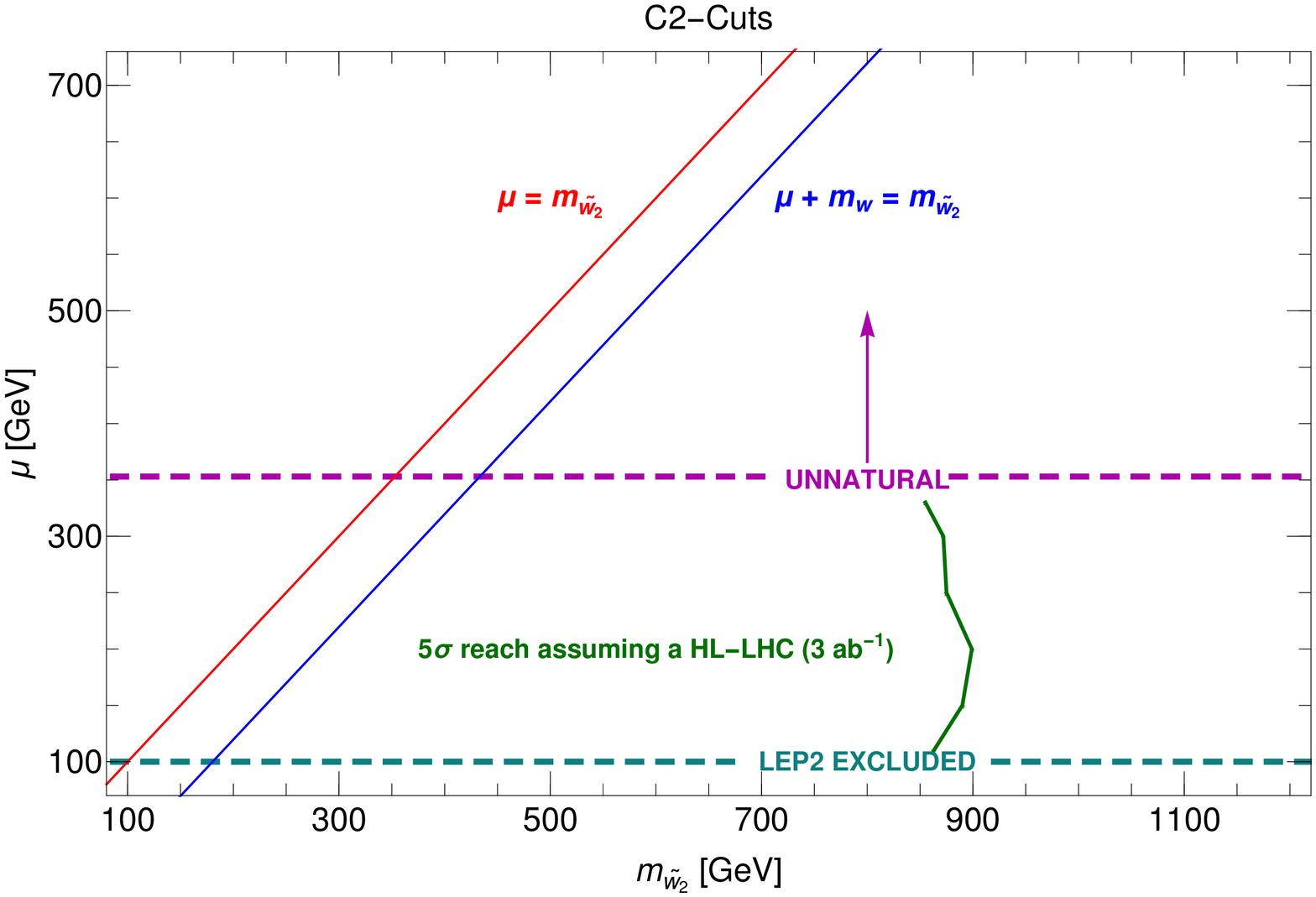}
\caption{ The left-hand frame shows the $5\sigma$ reach in the NUHM2
  model at the LHC and its luminosity upgrade for the monojet plus soft
  dilepton (labelled $\tz_1\tz_2 j$) and the same sign diboson $W^\pm
  W^\pm$ (labelled SSdB) channels discussed earlier in the text. Also
  shown are contours of several values of $\delew$. The green contour in
  the right-hand frame shows the reach of the HL-LHC via the same sign
  diboson channel from a different analysis (see text).}

\label{fig:nuhm2}
\end{center}
\end{figure}

We have seen that while there still is a potential for a SUSY discovery
in several channels, a signal is not guaranteed even at the luminosity
upgrade of the LHC. In models with gaugino mass unification, the
mono-jet plus soft dilepton channel and the same sign $WW +\eslt$
channels are the most promising in that they appear to cover the largest
portions of the parameter space with $\delew < 30$.  The situation,
within the NUHM2 framework, is summarized in the left frame of
Fig.~\ref{fig:nuhm2} from which we see that the mono-jet plus soft
dilepton yields an observable $5\sigma$ signal for $\mu \alt 250$~GeV at
the high luminosity LHC, while the same-sign $WW$ signal covers the
region with $m_{1/2} < 1.2$~TeV. (The corresponding gluino reach in
$m_{1/2}$ is slightly smaller.)  More interestingly, we see that with an
integrated luminosity of 3~ab$^{-1}$, LHC experiments should be
sensitive to the entire region of the $\delew < 30$ portion of the NUHM2
parmeter space! This exciting conclusion led to a reassessment the same
sign $WW$ signal in Ref.\cite{aspects} using madgraph/Pythia/Delphes
instead of ISAJET for the analysis. The corresponding reach, shown in
the right hand frame of Fig.~\ref{fig:nuhm2}, is about 10\% smaller than
that in the left hand frame.\footnote{This difference may be regarded as
  indicative of the systematic uncertainty in the reach projection.} but
the qualitative picture remains unaltered. At least in models with
gaugino mass unification (where $\delew < 30$ implies that the
neutralino mass gap is larger than $\sim 10$~GeV), the luminosity
upgrade of the LHC should be able to discover natural SUSY over most of
the parameter space via a signal in one (or both) of these channels.

Unfortunately, this optimistic conclusion may not carry over to models
with non-universal gaugino masses where electroweak gaugino masses can
be large (relative to $m_{\tg}$) without jeopardizing naturalness. This
has two effects. First, the $W^\pm W^\pm$ signal from wino pair
production may well be kinematically inaccessible. Second, larger values
of $M_{1,2}$ for fixed $\mu$ allows a higgsino mass splitting as small
as 3-4~GeV. The smaller mass gap implies softer leptons, and a
correspondingly  reduced efficiency for detecting the
dileptons in mono-jet events. Recent ATLAS projections
\cite{atlasdilep3} for the high luminosity LHC suggest that it may be
possible to detect the monojet plus soft dilepton signal with a
$5\sigma$ significance even for $m_{\tz_2}-m_{\tz_1}$ as small as $\sim
5$~GeV if $\mu < 220$~GeV ($\mu< 350$~GeV for exclusion at
95\%CL).\footnote{We are not aware of corresponding CMS analysis for
  such small mass gaps.} These early reach projections, though they allow for
discovery even with small higgsino mass gaps, are uncomfortably close to the
edge of the parameter space of natural SUSY. We hope and 
expect that these studies will be further refined, and that more
definitive results will be obtained. Until then, it seems prudent to
investigate what might be possible at accelerators that are being
considered for construction in the future.

\subsection{Electron-Positron Colliders} 

Since light higgsinos are $SU(2)$ doublets, they have typical
electroweak couplings, and so {\em must be} copiously produced at
$e^+e^-$ colliders, unless their production is kinematically suppressed.
Indeed cross sections for higgsino pair production proceses are
comparable to the cross section
for muon pair production if higgsino production is not kinematically
suppressed. Moreover, the higgsino pair production rate, for
higgsinos with masses comparable to $m_h$ exceeds that for $Zh$
production, so that these facilities may well be higgsino factories in
addition to being Higgs boson factories.
%
%
%
An electron-positron linear collider with a centre-of-mass energy of
500~GeV (and  upgradeable to 1~TeV)
that is being envisioned for construction is thus an obvious facility
for definitive searches for natural SUSY.  The issue is whether,
in light of the small visible energy release in higgsino decays, it is
possible to extract the higgsino signal above SM backgrounds. These
dominantly come from two-photon-initiated processes because those $2\to
2$ SM reactions can be efficiently suppressed by a cut on the visible
energy in the event.

The higgsino signal was first examined in Ref.~\cite{ilc} where the authors
studied two cases, both at a centre-of-mass energy just above the
production threshold for charged higgsino pair production.  The more
difficult of these (which is what we discuss here) was chosen so that
$m_{\tw_1} \simeq m_{\tz_2}=158$~GeV, and a mass gap with the neutralino
of just $\sim 10$~GeV, close to the minimum in models with gaugino mass
unification.  The small mass gap severely limits the visible energy, and
in this sense represents the maximally challenging situation within
models with unified gaugino mass parameters.

The most promising signals come from $e^+e^-\to\tw_1(\to
\ell\nu\tz_1)\tw_1(\to q\bar{q}\tz_1)$ which leads to $n_{\ell}=1$,
$n_j=1$ or 2 plus $\eslt$ events, and from $e^+e^- \to \tz_1\tz_2(\to
\ell\ell\tz_1)$ (with 90\% electron beam polarization to reduce $WW$
background) processes.  SM backgrounds can be nearly eliminated using
judicious cuts on the visible energy (signal events are very soft),
$\eslt$ and transverse plane opening angles between leptons and/or
jets. The higgsino signal could be extracted $\sqrt{s}=340$~GeV, and
an integrated luminosity of just a few fb$^{-1}$. We refer the reader
to Ref.~\cite{ilc} for details.

This early analysis has recently been re-examined in Ref.\cite{fujii}
with full Geant 4 based simulation of the ILD detector concept not only
for the two cases studied in Ref.\cite{ilc}, but also for an nGMM model
case for higgsinos with masses $\sim 155$~GeV, and a  neutral
higgsino mass gap is just 4.4~GeV. We refer the interested reader to
this study which confirms that the higgsino signal should be readily
detectable, even for the rather small mass gaps that may be possible in
natural SUSY. We conclude that an electron-positron collider will be
able to detect higgsino-pair production nearly all the way to the
kinematic limit, and further, that an electron-positron collider with
$\sqrt{s}\simeq 600$~GeV will probe the entire parameter space with
$\delew \leq 30$.

Aside from discovery, the clean environment of electron-positron
collisions also allows for precise mass measurements. For example, even
in the difficult case considered in Ref.~\cite{ilc} as well as the nGMM
case studied in Ref.\cite{fujii}, assuming an integrated luminosity of
500~fb$^{-1}$ at $\sqrt{s}=500$~GeV, a fit to the invariant mass
distribution of dileptons in $\tz_1\tz_2$ events allows the
determination of the neutralino mass gap, $m_{\tz_2}-m_{\tz_1}$.  A
subsequent fit to the distribution of the total energy of the two
leptons then allows the extraction of individual neutralino masses with
a precision of 0.7\% [1\%] for the case in Ref.\cite{ilc} [the nGMM case
  in Ref.\cite{fujii}]. These mass determinations, together with cross
section measurements using polarized beams, point to the production of
light higgsinos as the underlying origin of these novel events, and so
suggest a natural origin of gauge and Higgs boson masses.

\subsection{Energy Upgrade of the LHC} \label{subsec:LHC27}

The recent European Strategy Study envisages the possibility of 16 Tesla
dipole magnets which would allow the energy of the LHC to be increased
to 27~TeV in the existing LEP/LHC tunnel. The anticipated integrated
luminosity is 15~ab$^{-1}$ \cite{fcchh}. The increased energy offers an
opportunity to search for the coloured gluinos and stops of natural SUSY
whose production, as we saw in Sec.~\ref{subsec:lhc}, is kinematically
limited at the LHC.  A potential advantage of this search (because it
does not rely on an examination of the soft decays products of the
higgsinos) is that it would be insenstive to the details of the degree
of compression of the higgsino spectrum which limits the LHC reach via
the monojet plus soft-dileptons channel, or of the wino mass which
limits the LHC reach in the $W^\pm W^\pm +\eslt$ channel. It is,
therefore, possible that with the higher centre-of-mass energy gluino
and stop searches may offers the best possibility of the discovery of
natural SUSY in a wide variety of models.

Prospects for gluino and stop detection at a 33~TeV $pp$ collider
\cite{lhc33} were first examined in Ref.~\cite{jamie33}. This analysis
was then re-adapted for the high energy LHC (HE-LHC) a 27~TeV $pp$
collider for an assumed integrated luminosity of 15~ab$^{-1}$, assuming
that the gluino decays into a top and a (possibly virtual) stop, that
the stop decays promptly to higgsinos via $\tst_1 \to t\tz_{1,2}$ or
$\tst_1 \to b\tw_1$, and that the higgsino decay products are
essentially invisible \cite{jamie27}. As discussed in
Sec.~\ref{subsubsec:gst} pair production of heavy gluinos will lead to
events with up to 4 hard bottom jets (not all of which will be tagged as
$b$-jets) and large $\eslt$, while stop pair production results in up to
two tagged $b$-jets and large $\eslt$. It is relatively straightforward
to separate the SUSY signal from SM backgrounds, which dominantly come
from $b\bar{b}Z$ and $t\bar{t}Z$ with subdominant contributions from
$t\bar{t}, t\bar{t}b\bar{b},t\bar{t}t\bar{t}, t\bar{t}h$ and single $t$
production, by requiring at least two (four) very hard jets, at least
two of which are tagged as $b$-jets, for the signal from stop- (gluino-)
pair production together with very hard $\eslt$ along with other
analysis cuts. We  refer the
interested reader to Ref.\cite{jamie27} for further details. It was
found that after judicious cuts the $5\sigma$ reach of HE-LHC extends to
5.5~TeV for gluinos, and to 3.16~TeV for stops.\footnote{An independent
  analysis in Ref.\cite{ismail} finds, assuming $\tg \to t\bar{t}\tz_1$
  and $\tst_1\to t\tz_1$, a somewhat smaller reach of 4.8~TeV (2.8~TeV)
  for gluinos (top squarks).} The corresponding 95\%CL exclusion regions
  for both these sparticles extend out by about an additional 400~GeV.

\begin{figure}[tbp]
\begin{center}
\includegraphics[height=0.35\textheight]{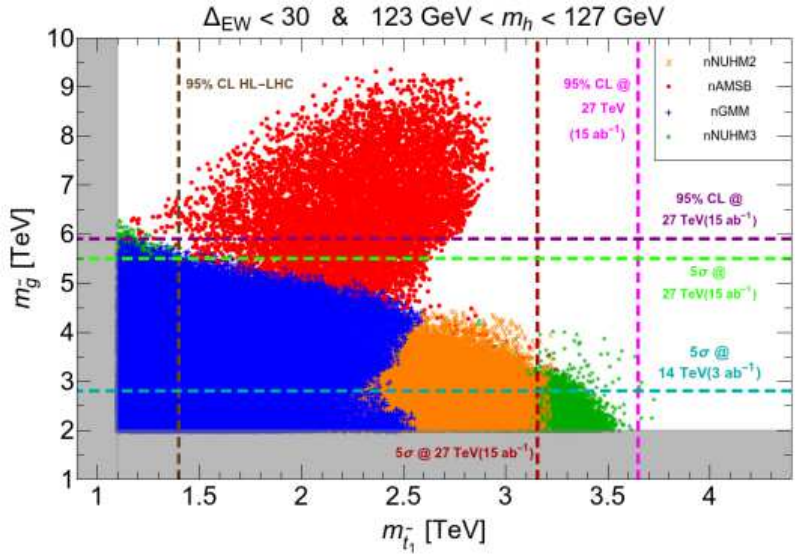}
\caption{ The gluino and stop reach of a $pp$ collider with
  $\sqrt{s}$=27~TeV, assuming an integrated luminosity of 15~ab$^{-1}$.
 Also shown is a scatter plot of points in the $m_{\tst_1}$ vs. $m_{\tz_1}$
  plane for various natural SUSY models with $\delew < 30$ 
introduced in the text: specifically, nNUHM2, nNUHM3, nGMM and nAMSB
models.
\label{fig:lhc27}}
\end{center}
\end{figure}

This is illustrated in Fig.~\ref{fig:lhc27} where the gluino and stop
reaches are shown in the $m_{\tst_1}-m_{\tg}$ plane by the horizontal
and vertical lines, respectively. Also shown are scatter plots of stop
and gluino masses in the various natural models with $\delew < 30$
introduced in Sec.~\ref{sec:spectra}: nNUHM2 (yellow crosses), nNUHM3
(green stars), nAMSB (red dots), and nGMM (blue pluses).  It is easy
to see that in all these natural models, there will be an observable
$5\sigma$ signal in at least one of the gluino or stop channels, and
for most of the parameter space, in both channels. Natural SUSY,
conservatively defined by no worse than 3\% electroweak fine-tuning,
would not evade detection at a 27~TeV $pp$ collider with an integrated
luminosity of 15~ab$^{-1}$. There may also be additional confirmatory
signals in other channels, but the observability of these signals
cannot be guaranteed. The discovery of stops and/or gluinos would
provide impetus for the construction of a yet higher energy collider
to snare the rest of the SUSY spectrum.

\subsection{Low Energy Measurements}

Precision measurements of SM particle properties offer an independent
avenue for probing new physics. This is not, however, the case for
the scenario that we have outlined, where our assumption that
sfermion mass parameters are very large
precludes the possibility of sizeable deviations from SM expectations
in processes such as $b\to s\gamma$, $b\to s\ell^+\ell^-$ or other
flavour violating processes, whose observed values are known to be 
compatible with SM predictions \cite{pdg}. We stress
that this assumption is not required by naturalness considerations, but
made to avoid unwanted flavour-changing-neutral currents. However, if the SM
computation of $(g_\mu-2)$ holds up to scrutiny and the 
measured value of the muon anomalous magnetic moment
\cite{gminus2} continues to deviate from its expectation in the SM
\cite{muontheory}, it will have to be due phenomena outside the class
of natural SUSY models that we find most promising.
 
Finally, we note that though SUSY contributions to the rate
for the exclusive rare decay $B_s\to \mu^+\mu^-$ do not decouple with the
super-partner mass scale, these are strongly suppressed for large values
of $m_A$. This is not a problem because for moderate to large values of
$\tan\beta$, $m_A^2 \simeq m_{H_d}^2 - m_{H_u}^2$ at tree level. Since
$m_{H_d}^2$ can be in the multi-TeV range without jeopardizing
electroweak fine-tuning (because the contribution of the $m_{H_d}^2$
term in Eq.~(\ref{eq:mZsSig}) is suppressed by the $(\tan^2\beta-1)$
factor), multi-TeV values of $m_A$ are typical in natural SUSY. This is
a plus because the measured value \cite{bsmmmeas} for the branching
fraction for this process is also in good agreement with the SM
prediction \cite{bsmmtheory}.

\subsection{Dark Matter} 

Since the LSP is expected to be higgsino-like and not far above the weak
scale in the simplest models with natural supersymmetry, it will
(co)annihilate rapidly to gauge bosons (via its large coupling to the $Z$
boson, and also via $t$-channel higgsino exchange processes) in the
early universe. This means that the measured cold dark matter density
{\em cannot arise solely from thermally produced higgsinos} in standard
Big Bang cosmology. Dark matter is thus likely to be multi-component. It
is important to note that naturalness considerations also impose
an upper bound on wino massses.
This, in turn, implies a lower limit on the gaugino content of the
higgsino-like LSP, and correspondingly on the neutralino-nucleon
scattering cross section which dominantly arises via $h$ exchange.
Indeed, it is then expected \cite{minilandBaer} that even with the
suppressed density, the XENONnT and LZ detectors \cite{xennT} will be
sensitive to the thermal higgsino signal from spin-independent
neutralino-nucleon scattering.\footnote{We remind the reader that there
  are the usual caveats to this conclusion. If physics in the sector
  that makes up the remainder of the dark matter entails late decays
  that produce SM particles, the neutralino relic density today could be
  further diluted, reducing the signal; see {\it e.g.}
  Ref.~\cite{howie_axion}. On the other hand, late decays of any associated
  saxion, axino or even string-moduli fields to the neutralino could
  enhance the neutralino relic density from its thermal value. The
  important lesson is that while the thermal relic density is
  interesting to examine, it would be imprudent to categorically exclude
  a new physics scenario based on relic density considerations alone,
  because the predicted relic density can be altered by the unknown
  (and, perhaps, unknowable) history of the Early Universe
  \cite{graciela}.} In models with gaugino mass unification, the upper
bound on $m_{\tg}$ leads to an even more stringent upper bound on the wino
mass, and the thermal higgsino signal would be detectable even at
XENON1T with its expected sensitivity to nucleon-neutralino cross
section at the $10^{-47}$~cm$^2$ level \cite{bbm}.

\section{Concluding Remarks} 
\label{sec:concl}

Weak scale supersymmetry stabilizes the electroweak scale and, in our
view, offers the best solution to the big hierarchy problem.  A
discovery of super-partners would mark a paradigm shift in particle
physics and cosmology. The non-observation of superpartners at LHC has
led some to express reservations about this far-reaching idea. As we
have discussed in Sec.~\ref{sec:scale} this is because the possibility
that the underlying SSB parameters of the underlying
theory might be correlated has been completely ignored. We recognize
that a credible high scale model of SUSY breaking that {\em predicts}
appropriate correlations among the SSB parameters and so automatically
has a modest degree of fine-tuning has not yet emerged, but we cannot
expect this until we understand the underlying SUSY breaking
mechanism.

To allow for these presently unknown parameter correlations, we advocate
using $\delew$, the electroweak fine-tuning measure for any discussion
of fine-tuning in SUSY. In this paper we consider models with $\delew >
30$ as definitely fine-tuned, and regard models that yield spectra with
$\delew < 30$ as possibly arising from an underlying theory with
moderate fine-tuning. We have checked that viable natural spectra exist
without a need for weak scale new particles beyond the MSSM, and have
argued that light higgsinos are the most robust consequence of SUSY
naturalness.

As discussed in Sec.\ref{sec:phen}, models with light higgsinos
potentially yield novel signals for supersymmetry at the LHC, the most
promising of which is the mono-jet plus soft-dilepton signal from
electroweak higgsino production with a radiation of a very hard QCD
jet. It appears that this signal will be observable at the luminosity
upgrade of the LHC with a significance $\ge 5\sigma$ over most of the
natural SUSY parameter space in models where gaugino mass unification is
assumed because the mass gap $m_{\tz_2}-m_{\tz_1}$ is then at least
10~GeV. In natural SUSY models where gaugino mass unification does not
hold, this mass gap may be as small as 4-5~GeV, so that the leptons from
$\tz_2$ decay tend to be softer and so more difficult to detect.  We are
excited by the early analysis by the ATLAS collaboration which suggests
that it may be possible to probe higgsinos via this channel even for
mass gaps substantially below 10~GeV. We urge our experimental
colleagues to continue to push this analysis to include the softest
leptons that they can as this will probe models with small mass
gaps. The stakes are high!

Also very interesting are $VV$, ($V=W^\pm ,Z$) $Vh$ and $hh+\eslt$ signals
from wino pair production, but these are not guaranteed because wino pair
production is kinematically limited by the energy of the
LHC. Nevertheless, if the signals turn out to be observable, the
relative strengths in the various channels could point to SUSY with
light higgsinos.

If gluinos and winos are too heavy to be accessible at the LHC, and the
neutralino mass gap is too small for the monojet plus soft dilepton
signal to be observable, we would need new facilities to detect natural
SUSY. One possibility is a linear electron-positron collider. We can
interpolate from the left frame of Fig.~\ref{fig:nuhm2} that a linear
collider operating at about 600~GeV would suffice to detect the
higgsinos of natural SUSY. Very interestingly, at the HE-LHC (a 27~TeV,
$pp$ collider expected to accumulate an integrated luminosity of
15~ab$^{-1}$) that is being considered for construction in the existing
LEP/LHC tunnel, at least one of the gluino or the stop of natural SUSY
(and likely both over most of the natural SUSY parameter space) should
be detectable with a significance $\ge 5\sigma$, independent of the
details of the electroweak-ino spectra. Natural SUSY, as we have defined
it, would not escape detection at such a facility.
 
Before closing, we note that in advocating the use of $\delew$ for
discussions of fine-tuning, we have adopted a bottom-up approach to
naturalness. Baer and his collaborators \cite{stringy} have recently
analysed SUSY naturalness from the top-down perspective of the string
landscape, arguing that one value of an observable is more natural
than another if the number of phenomenologically acceptable string
vacua that lead to this value is larger. With some assumptions about
the distribution of SUSY breaking $F$- and $D$-terms in these vacua,
they find that the number of vacua grows with the SUSY breaking scale,
favouring large values of SSB terms. However, in order
to obtain a universe with the diversity of nuclei that we observe, one
has to require (assuming everything is kept fixed) that the weak scale
is not far from its phenomenological value \cite{agrawal}. The
universe that we live in is then the result of a delicate balance
between these two (somewhat opposing) requirements. The
authors of Ref.\cite{stringy} conclude that the anthropic requirement
that the weak scale be within about a factor four of its observed
value, with $|\mu|$ not much larger than the weak scale, leads to low
energy SUSY models with $\delew < 30$, and first/second generation
sfermion masses in the ten TeV range. These are exactly the
characteristics of the models that we have discussed in our bottom-up
approach! A detailed discussion of these speculative landscape ideas is
beyond the scope of this paper, and we will refer the interested
reader to a companion article \cite{bspaper} in this Volume.

In summary, SUSY GUTs pioneered by many authors during the 1980s remain as
promising as ever. Moreover, the original aspirations of early workers
on weak scale supersymmetry outlined in Sec.~\ref{sec:intro} remain
unchanged, if we accept that
\bi
\item ``accidental cancellations'' at the
few percent level are ubiquitous and may not require explanation, and
\item dark matter may be multi-component.\footnote{Given that visible
  matter which comprises a small mass fraction of the total matter
  content already consists of several components, this is hardly a
  stretch.}  \ei The fact that low scale physics is only
  logarithmically (and not quadratically) sensitive to the scale of
  ultra-violet physics remains a very attractive feature of softly
  broken SUSY models, and leads to an elegant resolution of the big
  hierarchy problem.  That it is possible to find phenomenologically
  viable models with low electroweak fine-tuning leads us to speculate
  that our understanding of UV physics is incomplete, and that there
  might be high scale models with the required parameter correlations
  that will lead to comparably low values of the true fine-tuning
  parameter $\delbg$. The supergravity GUT paradigm remains very
  attractive despite the absence of sparticle signals at the LHC. We
  urge the continued exploration of the energy frontier at the
  HL-LHC, at future electron-positron colliders with $\sqrt{s} \agt
  600$~GeV, or at the proposed energy upgrade of the LHC where it will be
  possible to definitively test the ideas reviewed here.

\section*{Acknowledgments}
I am grateful to H.~Baer, V.~Barger, J.~Gainer, P.~Huang, D.~Mickelson,
A.~Mustafayev, M.~Padefkke-Kirkland, M.~Savoy, D.~Sengupta, H.~Serce,
W.~Sreethawong and T. Ghosh for discussions and collaboration on much
of the work described here.

%

\end{document}